\documentclass[1p,times,procedia]{elsarticle}
\usepackage{nupha_ecrc}

\volume{00}

\firstpage{1}

\journalname{Nuclear Physics A}

\runauth{Chun Shen}

\jid{nupha}

\jnltitlelogo{Nuclear Physics A}

\usepackage{amssymb}

\usepackage[figuresright]{rotating}

\begin{document}

\begin{frontmatter}



\dochead{The XXVth International Conference on Ultrarelativistic Nucleus-Nucleus Collisions}

\title{Electromagnetic Radiation from QCD Matter: Theory Overview}


\author{Chun Shen}

\address{Department of Physics, McGill University, 3600 University Street, Montreal, QC, H3A 2T8, Canada}

\begin{abstract}
Recent theory developments in electromagnetic radiation from relativistic heavy-ion collisions are reviewed. Electromagnetic observables can serve as a thermometer, a viscometer, and tomographic probes to the collision system. The current status of the ``direct photon flow puzzle'' is highlighted.


\end{abstract}


\end{frontmatter}


\section{Introduction}
\label{intro}

Relativistic heavy-ion collision experiments conducted at the Relativistic Heavy-Ion Collider (RHIC) and the Large Hadron Collider (LHC) create an environment at a temperature of a trillion degrees, to study the property of dense nuclear matter. Electromagnetic probes, such as direct photons and dileptons ($e^+e^-$ and $\mu^+\mu^-$ pairs) are recognized as valuable messengers in such collisions. Because photons and dileptons interact only electromagnetically, they are able to penetrate the medium and carry almost undistorted dynamical information and report on conditions existing at their production point. Such probes are sensitive to the early stages of the collision system, to thermal and transport properties of the quark-gluon plasma (QGP), and to the dynamical evolution proceeding from the cross-over regions to the hadronic phase. The recent direct photon measurements show large yields of photons and large momentum anisotropies for $p_T < 4$\,GeV\,\cite{Adare:2014fwh,Adam:2015lda,Adare:2015lcd,Chatterjee:2005de}. Theoretical calculations still underestimate these challenging measurements: this tension has been dubbed the ``direct photon flow puzzle''.

In this proceeding, I review recent theoretical developments which demonstrate that electromagnetic probes can provide us with   important information, complementary to that carried by the majority of hadronic observables. The progress in theory towards resolving the ``direct photon flow puzzle'' is also discussed. 

\section{Thermometer}

Real and virtual photons are regarded as useful tools for experimentally accessing the temperature of the QGP created in heavy-ion collisions\,\cite{Shuryak:1978ij,Hwa:1985xg}. The slopes of the photon and dilepton spectra encode temperature information of the collision system.  

\begin{figure*}[h!]
\centering
\begin{tabular}{cc}
  \includegraphics[width=0.45\linewidth]{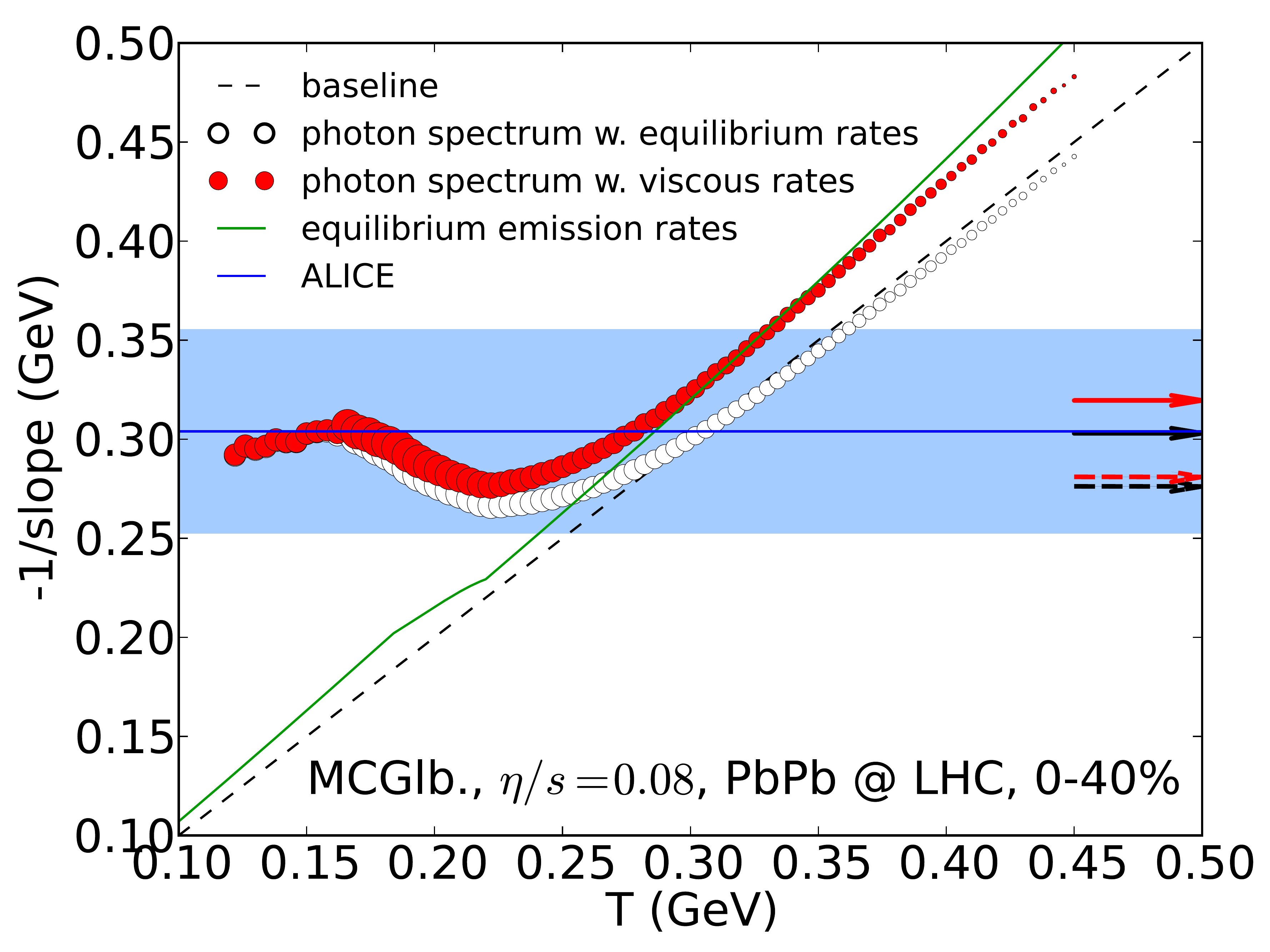} & 
  \includegraphics[width=0.45\linewidth]{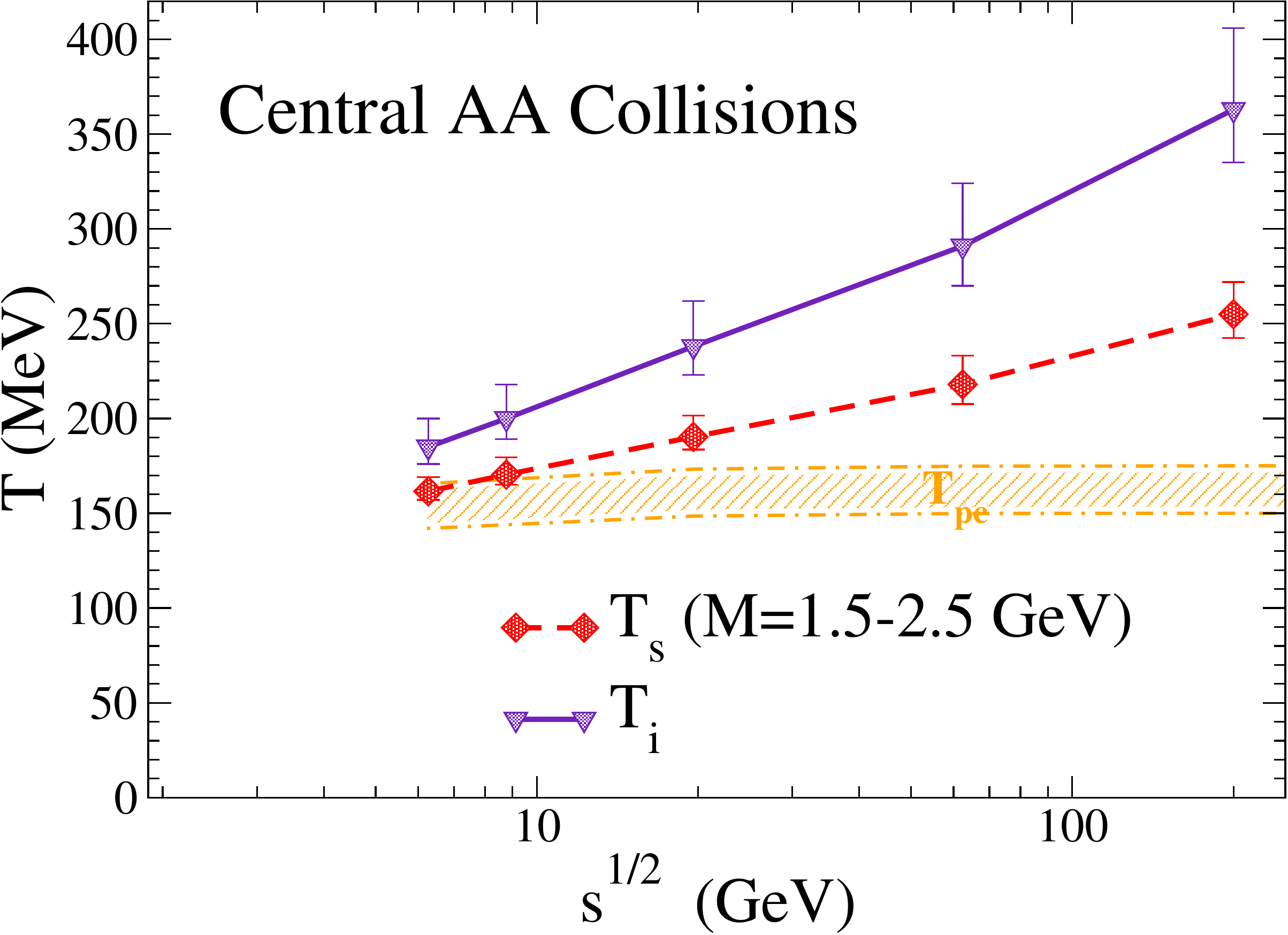}
\end{tabular}
\caption{{\it Left Panel:} The inverse slope of thermal photon spectrum emitted from hydrodynamic fluid cells as a function of the local temperature. The plot is taken from Ref.\,\cite{Shen:2013vja}. {\it Right Panel:} Temperatures ($T_s$) extracted from the slope of thermal dilepton invariant mass spectrum in the range $M = 1.5 - 2.5$\,GeV, compared with the initial temperature of the fireball $T_i$ at the different collision energies. The figure is taken from Ref.\,\cite{Rapp:2014hha}. }
\label{fig1}
\end{figure*}

Direct photon spectra have been measured by the PHENIX, STAR, and ALICE Collaborations in heavy-ion collisions at the top RHIC and LHC energies\,\cite{Adare:2014fwh,Adam:2015lda}. The low $p_T$ part of the spectra can be well characterized by their inverse logarithmic slope $T_\mathrm{eff}$. The PHENIX Collaboration reported $T_\mathrm{eff} = (239 \pm 25^\mathrm{stat} \pm 7^\mathrm{sys})$\,MeV for 0-20\% Au+Au collisions at $\sqrt{s_\mathrm{NN}} = 200$\,GeV \cite{Adare:2014fwh} and the ALICE Collaboration found $T_\mathrm{eff} = (304 \pm 11^\mathrm{stat} \pm 40^\mathrm{sys})$\,MeV in 0-20\% Pb+Pb collisions at $\sqrt{s_\mathrm{NN}} = 2.76$\,TeV \cite{Adam:2015lda}. Quantitive  studies \cite{vanHees:2011vb, Shen:2013vja} have shown that thermal photons emitted from $T < 250$\,MeV received a significant blue-shift from hydrodynamic flow, as illustrated in the left panel of Fig.\,\ref{fig1}. 
%
%
A solid extraction of the initial temperature of the collision system requires detailed comparisons between experimental data and dynamical model simulations. 

On the other hand, thermal dilepton invariant-mass spectra are free from the blue-shift contamination. The authors in Ref.\,\cite{Rapp:2014hha} demonstrated that the slope of dilepton spectrum in the intermediate mass region (IMR), $1.5$\,GeV\,$< M < 2.5$\,GeV, could provide clean temperature information about the collision system. The extracted averaged temperature was about 30\% lower than the corresponding initial one at the top RHIC energy (see right panel of Fig.\,\ref{fig1}). This difference shrank as the collision energy decreased. 
Thermal dilepton invariant mass spectra provide model-independent information about the collision system.
In addition, Refs.\,\cite{Rapp:2014hha,Rapp:2013ema,Bratkovskaya:2011wp,Vujanovic:2015gba} showed that dilepton invariant mass spectra were valuable tools to probe the properties of the baryon-rich fireball in the RHIC Beam Energy Scan (BES) program \cite{Xu:2014jsa,Adamczyk:2015lme}. Finally, dilepton invariant mass spectrum can help us to study the in-medium modification of the vector meson ($\rho$-meson) spectral function, which has a direct connection to chiral restoration during the phase transition\,\cite{Hohler:2015iba}. 
\section{Viscometer}
In relativistic heavy-ion collisions, the medium does not stay in thermal equilibrium during its evolution. Out-of-equilibrium dynamics can therefore influence the electromagnetic observables. 

Effects on thermal photon emission owing to locally anisotropic particle distributions were first investigated in Ref.\,\cite{Schenke:2006yp,Dusling:2009bc}. More complete calculations of the shear viscous corrections to the photon production rates were performed in Ref.\,\cite{Dion:2011pp,Shen:2014nfa,Shen:2014thesis}, and the related phenomenological impacts were studied in Ref.\,\cite{Dion:2011pp,Shen:2013cca,Shen:2014cga}. The viscous hydrodynamic evolution by itself was found to increase the net photon elliptic flow\,\cite{Shen:2014cga}, because the initial temperature of the system is lower with non-zero specific shear viscosity in order to compensate for the entropy production during the evolution. This reduced early stage QGP photon emission at high $p_T$, which then increased the relative weight of photons from later stages which carry a large momentum anisotropy. Similar effects were also found in Ref.\,\cite{Bhattacharya:2015ada}. The further inclusion of the viscous corrections to the photon emission rates reduced the thermal photon $v_n$\,\cite{Dion:2011pp,Shen:2013cca}. 

Effects from bulk viscosity were recently studied in Ref.\,\cite{Ryu:2015vwa,Paquet:2015lta}. The inclusion of a non-vanishing bulk viscosity near the phase transition was recently found to be essential to provide a good description of identified hadron mean-$p_T$ measurements\,\cite{Ryu:2015vwa}. The extra entropy production from bulk viscosity increases the space-time volume in the late hadronic phase by about 50\%, which allows more thermal photon radiation. Another consequence of the inclusion of bulk viscosity is to reduce the hydrodynamical radial flow by $\sim$10\% at the late stage of the evolution: this transport coefficient slows down the fireball expansion and weakens the blue shift of the thermal photon spectrum. Both effects together increase the thermal photon yields in the low $p_T$ regions and shift the peak of the direct photon $v_2(p_T)$ towards the low $p_T$ regions\,\cite{Paquet:2015lta}.

Finally, virtual photons, measured as lepton pairs, have also been shown to be a clean and sensitive probe of the out-of-equilibrium dynamics of the system. Recent studies  have demonstrated that, compared to hadronic observables, the thermal dilepton spectrum and its flow anisotropy show a larger sensitivity to the early time dynamics, to the system's shear stress tensor $\pi^{\mu\nu}$, to the temperature dependence of shear viscosity $\eta/s(T)$, and even to the choice of the second order transport coefficient $\tau_\pi$ \cite{Ryblewski:2015hea,Vujanovic:2014vwa}. 

\section{Status of resolving the ``direct photon flow puzzle''}
Ever since the unexpectedly large direct photon elliptic flow reported by the PHENIX Collaboration in Au+Au collisions at RHIC\,\cite{Adare:2011zr}, the tension with the experimental measurements has generated considerable theoretical effort to resolve the ``direct photon flow puzzle''.

\begin{figure*}[t!]
\centering
\begin{tabular}{ccc}
  \includegraphics[width=0.31\linewidth]{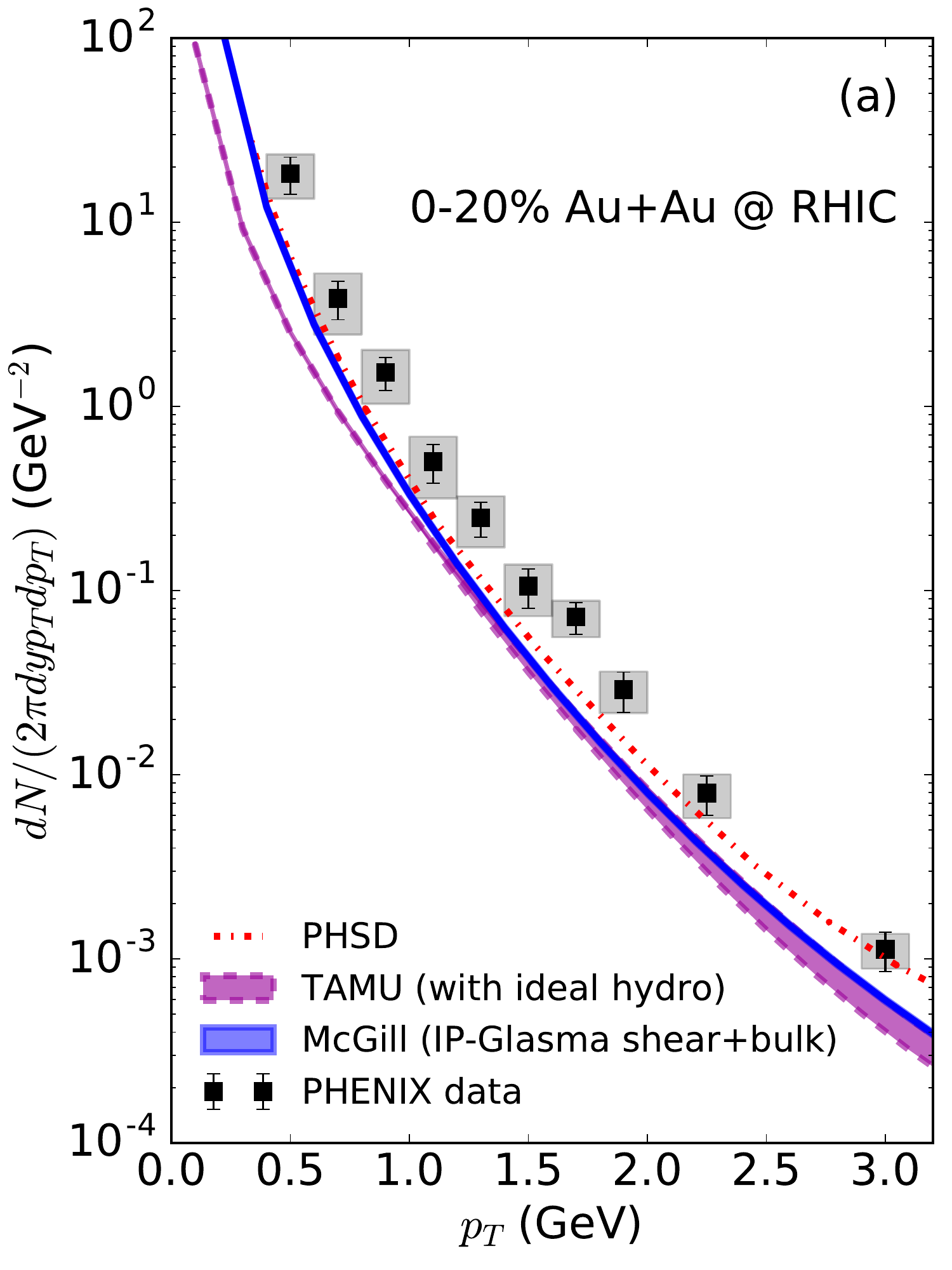} & 
  \includegraphics[width=0.31\linewidth]{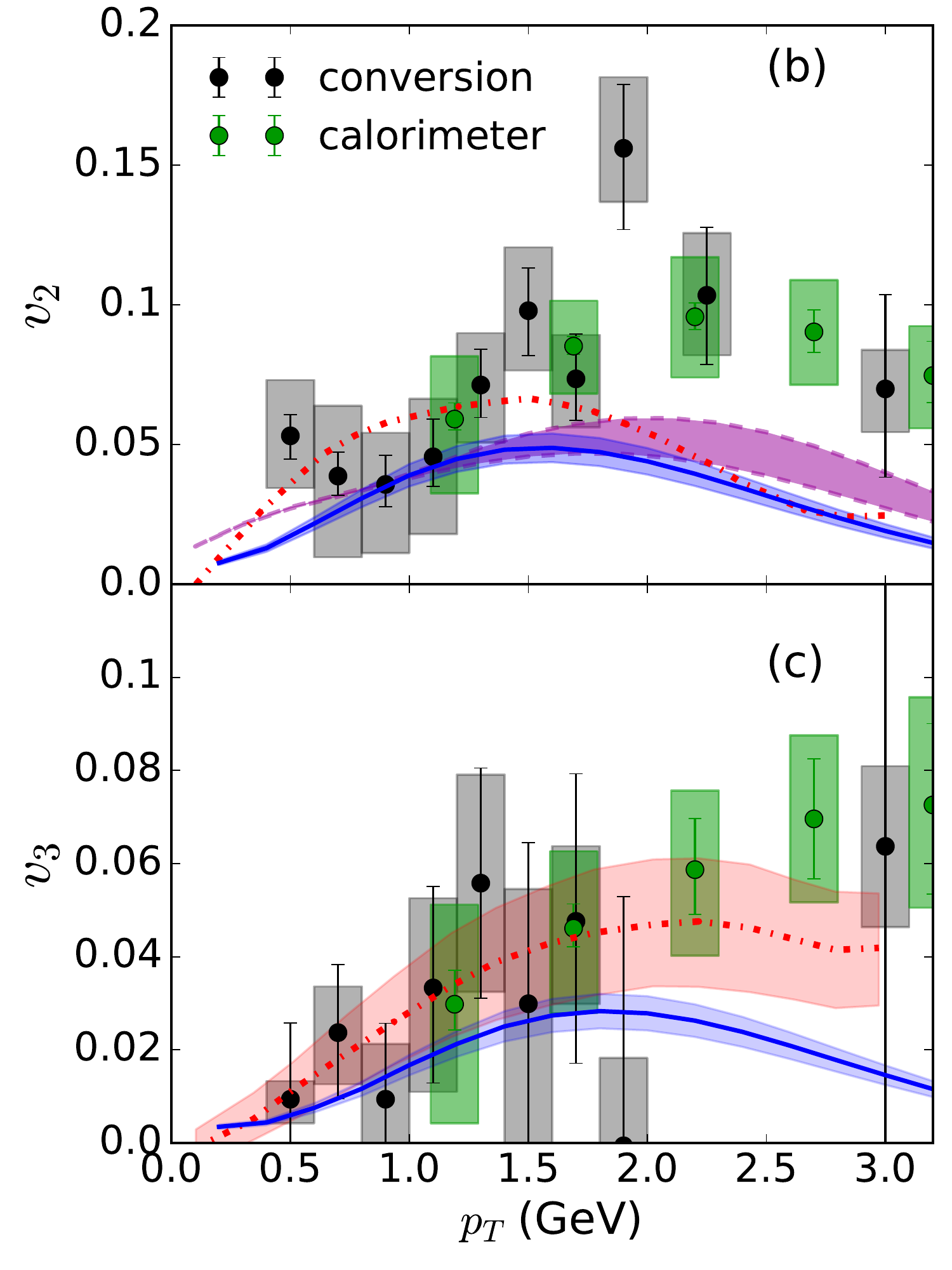} &
  \includegraphics[width=0.31\linewidth]{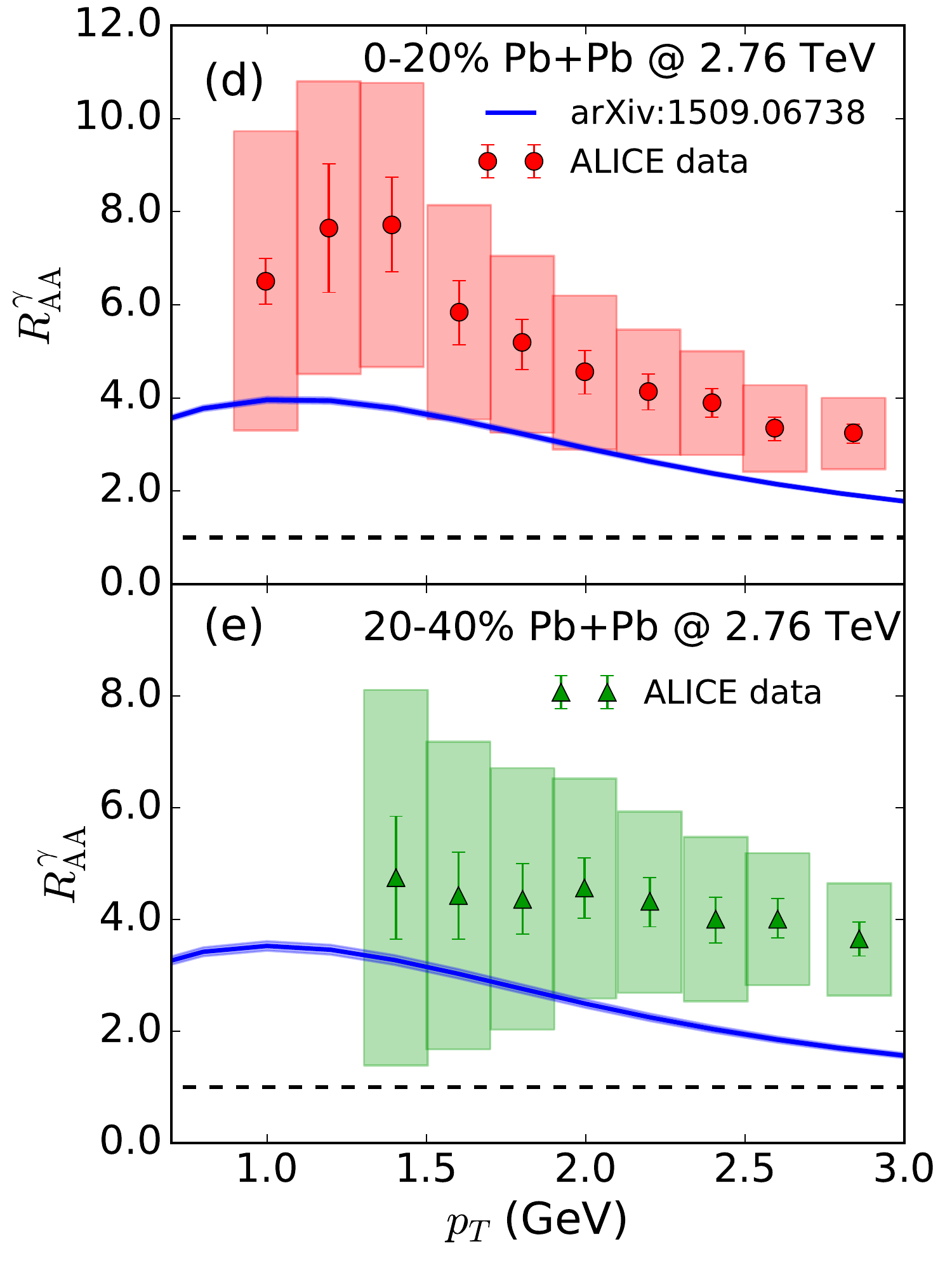}
\end{tabular}
\caption{{\it Panels (a)-(c):} Theoretical calculations of direct photon spectra and their flow anisotropy $v_2(p_T)$ and $v_3(p_T)$ compared with the PHENIX measurements in 0-20\% Au+Au collisions at $\sqrt{s_\mathrm{NN}}=200$\,GeV\,\cite{Adare:2014fwh,Adare:2015lcd}. The uncertainty band in the TAMU results presented two choices of the prompt sources\,\cite{vanHees:2014ida}. The PHSD results were taken from Ref.\,\cite{Linnyk:2015tha}. The error band is dominated by the uncertainty in the modeling of the cross sections. Statistical errors were indicated as a band in the McGill results\,\cite{Paquet:2015lta}. The legend in (a) applies to Panels (b) and (c). {\it Panels (d), (e):} The nuclear modification factor of direct photon $R^\gamma_\mathrm{AA}$ in 0-20\% and 20-40\% Pb+Pb collisions at $\sqrt{s_\mathrm{NN}} = 2.76$\,TeV \cite{Adam:2015lda,Sahlmueller:2015rhy}.}
\label{fig4}
\end{figure*}

Fig.\,\ref{fig4} captures the current situation of model-data comparisons for direct photon spectra and their anisotropic flow coefficients.\footnote{Please see Ref.\,\cite{Shen:2015nto} for a recent detailed review about the theoretical efforts to resolve the ``direct photon flow puzzle''.} With respect to some earlier calculations \cite{Chatterjee:2013naa,Shen:2014lpa}, the tension between experimental measurements and theory is now considerably reduced. Calculations with different hydrodynamic evolution   \cite{Paquet:2015lta,vanHees:2014ida} give  similar results for direct photon spectrum and elliptic flow coefficient for $p_T > 1$\,GeV. The major improvements common to these calculations were the inclusion of a more complete set of hadronic emission channels, namely the contributions from $\rho$-spectral function\,\cite{Rapp:1999ej,Turbide:2003si},   $\pi\pi$ bremsstrahlung\,\cite{Liu:2007zzw,Heffernan:2014mla}, $\pi \rho \omega$ channels\,\cite{Holt:2015cda},
and short-lived resonances feed-down\,\cite{vanHees:2014ida,Shen:2014lpa}.
The remaining difference below 1 
GeV originates from different extrapolations procedures for the prompt photon source. The McGill group extrapolated the pQCD prompt photon source to low $p_T$ using results from different fragmentation scales\,\cite{Paquet:2015lta}, while the TAMU group suppressed soft prompt photon production with formation times longer than the thermalization time of the collision system\,\cite{vanHees:2014ida,Turbide:2003si}. By considering event-by-event fluctuations as well as both shear and bulk viscous effects on the photon production mechanisms, the results presented in Ref.\,\cite{Paquet:2015lta} achieve good agreement with experimental data, and also an make statements on the triangular flow of direct photons. Finally, theoretical uncertainties in the photon emission related to nonperturbative physics were investigated in Ref.\,\cite{Paquet:2015lta}, by considering the Semi-QGP scenario \cite{Hidaka:2015ima,Satow:2015oha,Gale:2014dfa} and an alternative hadronic emission rate \cite{Dusling:2009ej} (also see discussions in Ref.\,\cite{Shen:2015nto}). Adopting those, the variation in the final result is smaller than current data systematic errors. More accurate measurements could however distinguish between those different techniques and therefore set constraints on photon emission rates. 

Considering alternatives to hydrodynamic approaches, direct photons were studied using  microscopic transport simulations\,\cite{Linnyk:2015tha}. The PHSD model provides a  level of agreement with the experimental measurement in 0-20\% Au+Au collisions shown in Fig.\,\ref{fig4}a. The centrality dependence of direct photon observables in semi-peripheral collisions were better reproduced in the PHSD model owing to larger hadronic bremsstrahlung contributions\,\cite{Linnyk:2015tha}. A channel-by-channel comparison between the transport and hydrodynamic approaches was presented in Ref.\,\cite{Shen:2015nto}. One found a similar QGP photon emission from the two approaches, but a rather different contribution from hadronic bremsstrahlung processes. The PHSD model additionally includes meson-baryon bremsstrahlung, using an improved soft photon approximation \cite{Linnyk:2015tha}. For $\pi\pi$ bremsstrahlung, the PHSD model produced $\sim$4 times more photons than in hydrodynamic calculations\,\cite{Shen:2015nto}. This might suggest additional sources from the dilute hadronic phase, which are not currently modelled in the hydrodynamic framework. A closer look at the space-time structure of hadronic photon emission will elucidate the origin(s) of the difference between the two models. Comparisons with recent photon calculations using a coarse-grained UrQMD medium \cite{Endres:2014zua,Endres:2015egk} in the hadronic phase will also shed light on this issue. 

\section{Event-by-event tomography}

\begin{figure*}[h!]
\centering
\begin{tabular}{ccc}
  \includegraphics[height=0.28\linewidth]{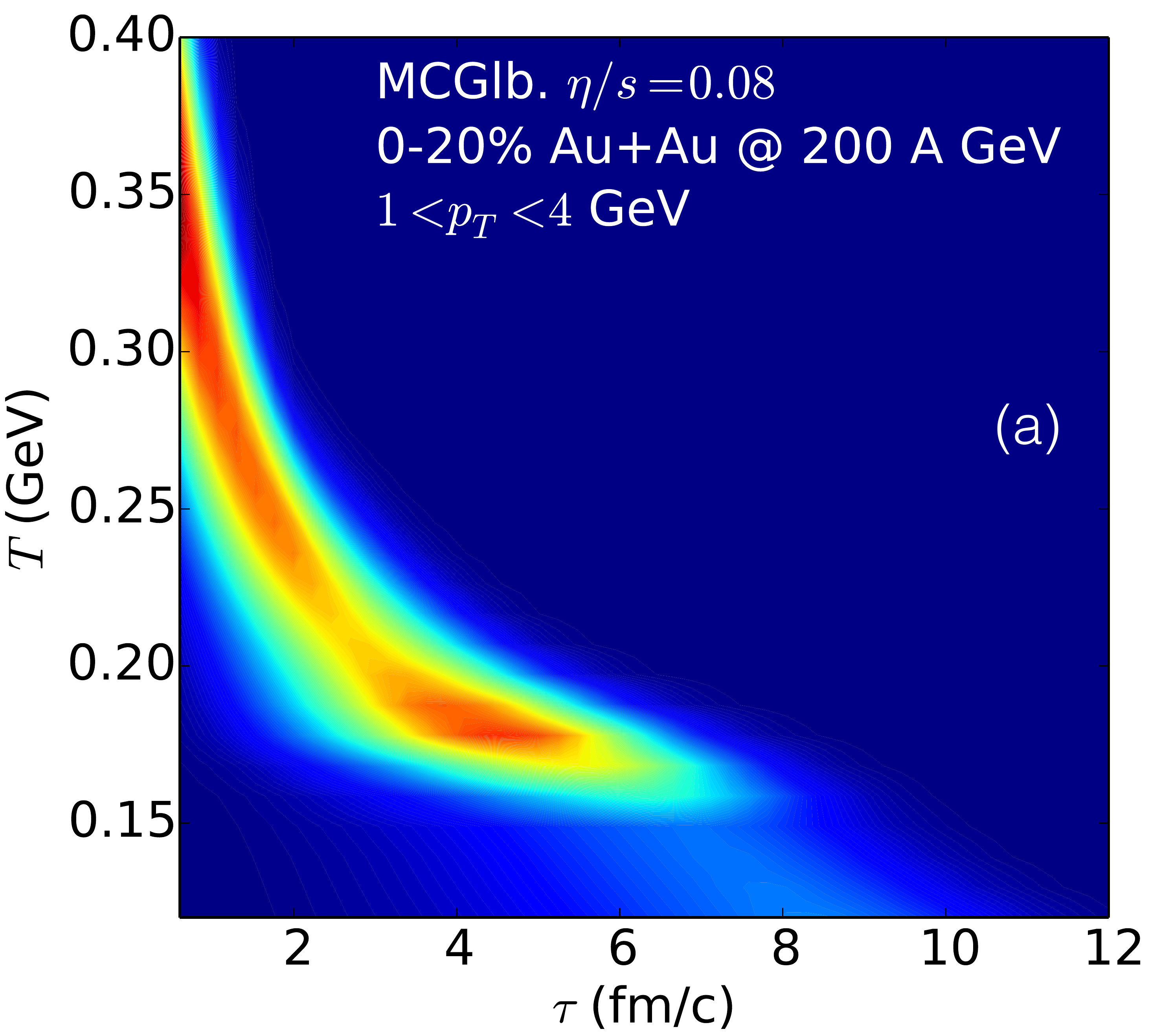} & 
  \includegraphics[height=0.28\linewidth]{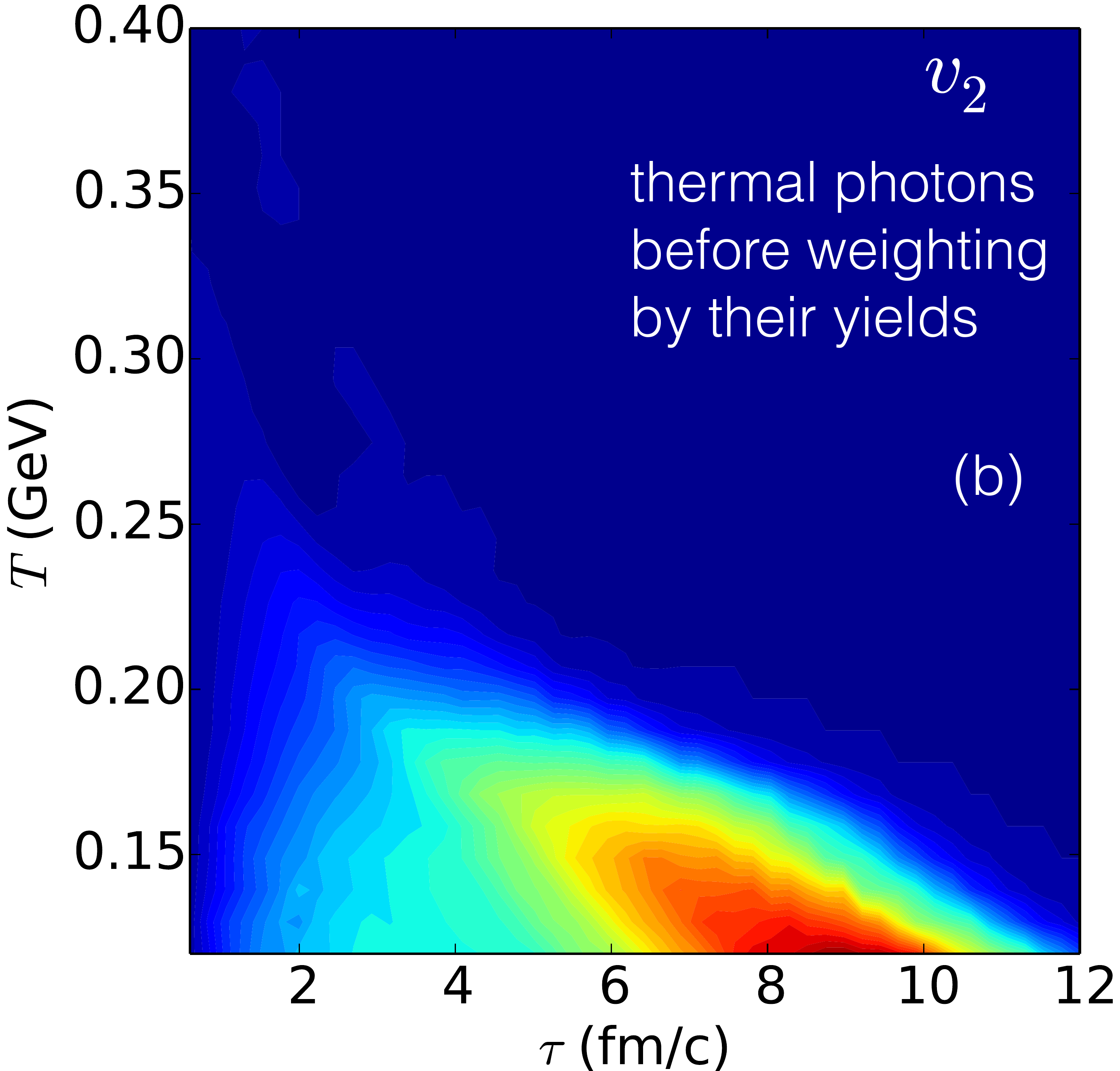} & 
  \includegraphics[height=0.28\linewidth]{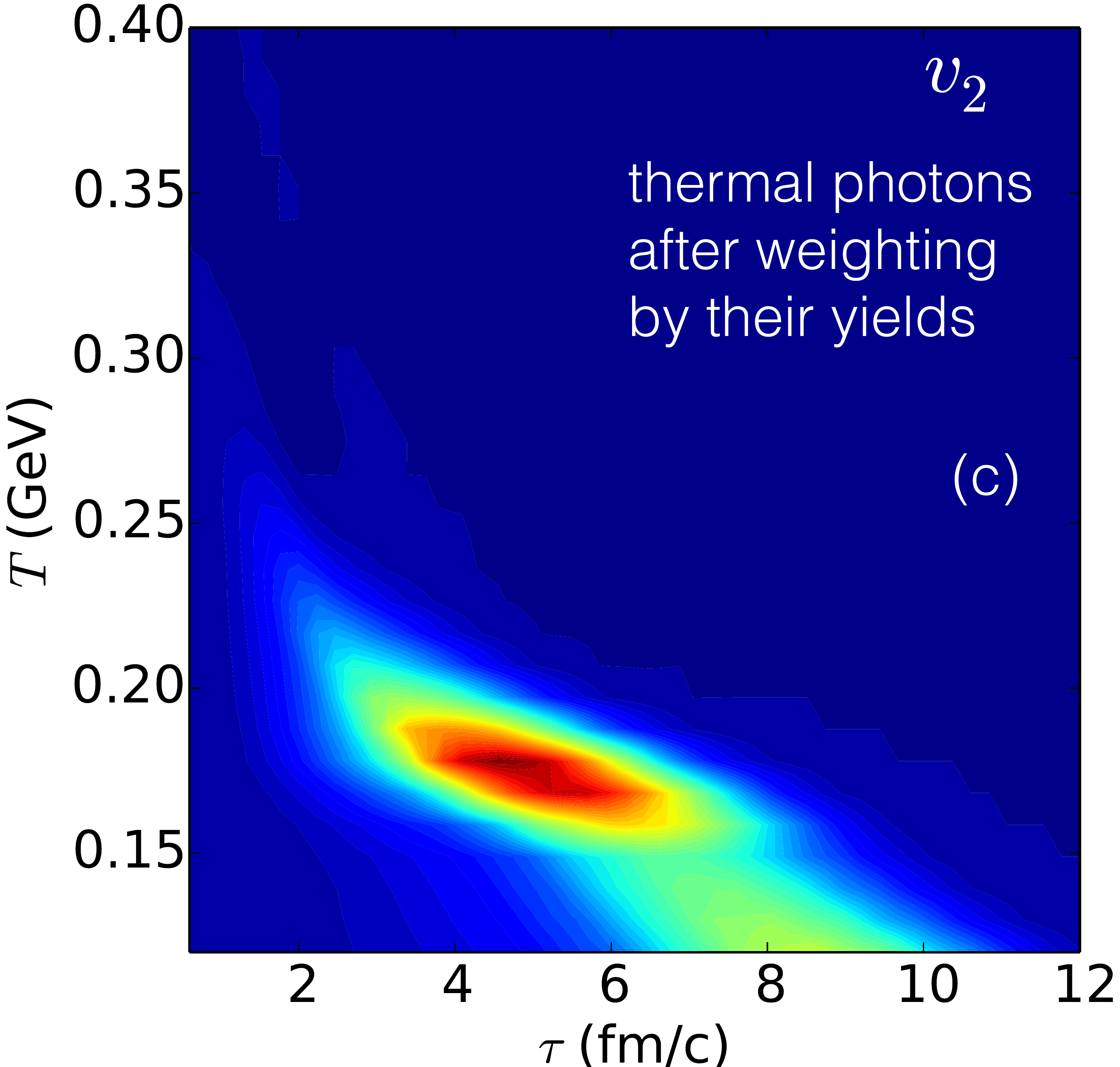}
\end{tabular}
\caption{Color contour plots for thermal photon yield {\it Panel (a)} and its elliptic flow coefficient {\it Panels (b) and (c)} with and without multiplicity weights as function of local temperature and proper time of the fluid cells. }
\label{fig3}
\end{figure*}

Thermal photons can also  serve as a tomographic probe which reports on the dynamics and property of the interior of the fireball, in analogy with the now common  medical CT scan.  A space-time analysis of thermal photon production in 0-20\% Au+Au collisions is shown in Fig.\,\ref{fig3}a.
%
%
A two-wave structure of thermal photon emission was found \cite{Shen:2013cca}. The first wave is from high temperature regions during the early stage of the evolution and the second wave sits near the cross-over region and is connected with a large space-time volume \cite{Shen:2013cca,vanHees:2014ida}.  
The development of thermal photon elliptic flow in every fluid cell is illustrated in Fig.\,\ref{fig3}b. Thermal photons emitted at the late stage of the collisions carry large elliptic flow. After weighting with the photon yield in every fluid cell, we observe that the most of the elliptic flow signal is coming from the cross-over region between $\tau = 3-8$ fm/$c$ shown in Fig.\,\ref{fig3}c. In contrast to charged hadrons, whose momentum distributions only freeze-out below $T_\mathrm{dec} = 120$\,MeV, thermal photons indeed carry direct dynamical information  from the higher temperature regions and from earlier time.

Furthermore, color tomographic probes, such as  energetic quarks, will lose some of their energy and radiate soft photons when penetrating the QGP medium. Photons from these jet-medium interactions were shown as an important source in the direct photon spectra for $2 < p_T < 4$\,GeV\,\cite{Qin:2009bk}. Recently, a new photon production process during the jet-medium interaction was computed in Ref.\,\cite{Qin:2014mya}. The investigation of its importance in phenomenological studies with realistic event-by-event hydrodynamic medium is an active and ongoing research subject. Additionally, photon production during the hadronization stage is proposed in Ref.\,\cite{Campbell:2015jga,Young:2015adw}. These processes probe different space-time regions of the evolving medium. Their relative importance in the final direct photon signal requires a quantitative calculation of their absolute yield, and a comparison with the dominant thermal sources. The current state-of-the-art hydrodynamic framework\,\cite{Paquet:2015lta,Shen:2014vra} will provide an excellent test ground for these new ideas.

\begin{figure*}[h!]
\centering
\begin{tabular}{cc}
  \includegraphics[width=0.4\linewidth]{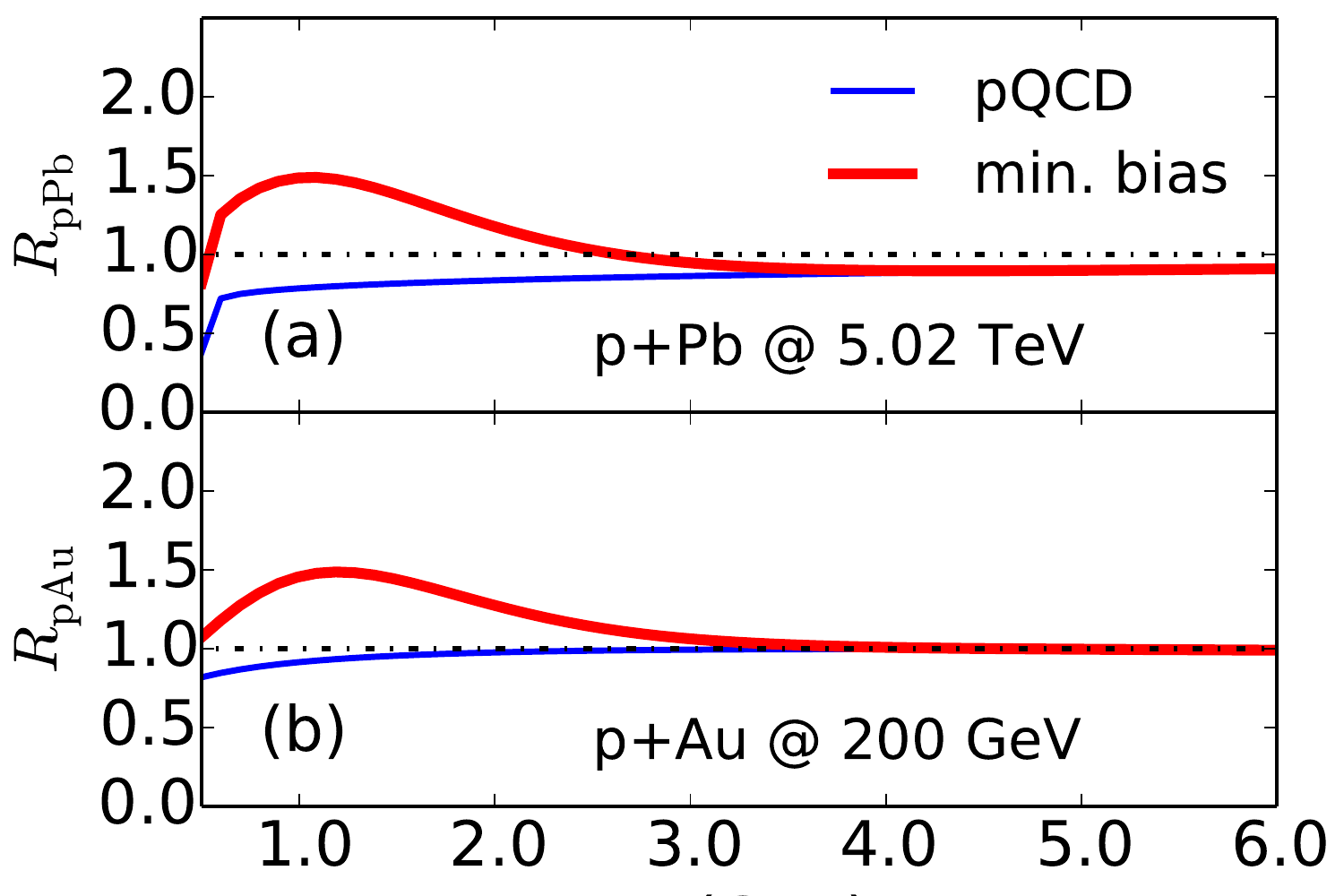} & 
  \includegraphics[width=0.4\linewidth]{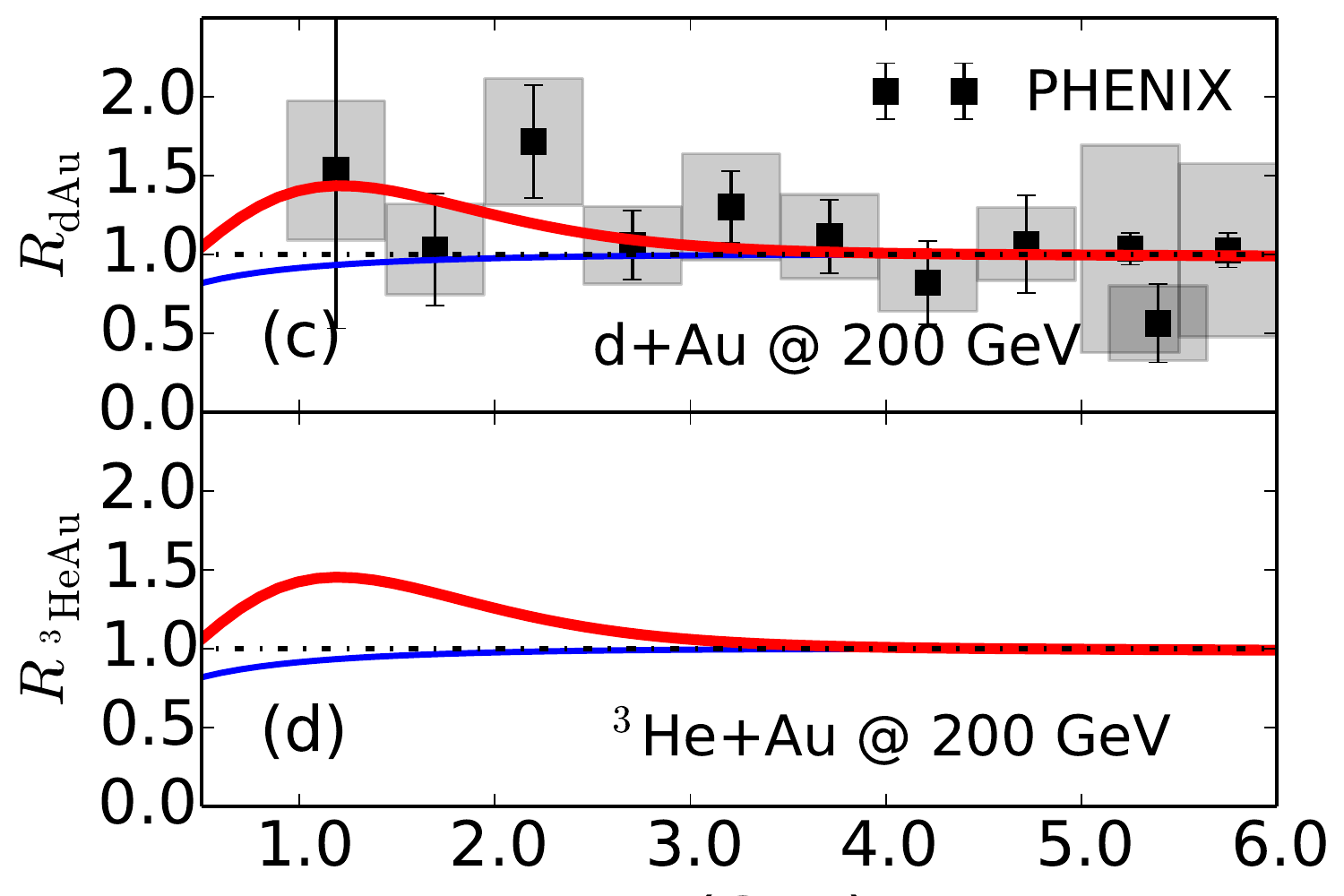} 
\end{tabular}
\caption{The nuclear modification factor $R^\gamma_\mathrm{AA}$ of direct photons in small collision systems\,\cite{Shen:2015qba}. Minimum bias calculation in $d$+Au collisions at 200 GeV is compared with the PHENIX measurement\,\cite{Adare:2012vn}.}
\label{fig5}
\end{figure*}

Finally, a recent work\,\cite{Shen:2015qba} studied the direct photon production in small collision systems at the RHIC and LHC energies. Fig.\,\ref{fig5}(a-d) presents theory predictions of the direct photon nuclear modification factor $R^\gamma_\mathrm{AA}$  for these cases. A $\sim$50\% thermal enhancement over the prompt pQCD contribution in the direct photon yield was found for $p_T < 3$\,GeV in  minimum bias collisions. Even though the overall strength of the signal is small when compared to the thermal enhancement in Pb+Pb collisions shown in Figs.\,\ref{fig4}d and \ref{fig4}e, such a measurement can serve to support the existence of a thermalized QGP in small systems at  RHIC and the LHC, therefore acting to complement the evidence carried by hadronic observables. The existing data in $d$+Au collisions at the top RHIC energy still carry large uncertainties and can this not be used to draw decisive conclusion on the existence of a thermal component\,\cite{Adare:2012vn}. Future electromagnetic measurements with improved accuracy and for different collision systems (p+Au, $^3$He+Au, and p+Pb) will guide our understanding of  the dynamics and of the properties of small QGP droplets. 

\section{Conclusion and outlook}

Modeling electromagnetic observables is  a double-edged sword, as it is sensitive to every aspect of the relativistic heavy-ion collisions:  out-of-equilibrium dynamics of the collision systems; thermal and transport properties of the QGP matter; and non-perturbative aspects near the cross-over region. For the very  same reasons, a well-calibrated calculation can turn the electromagnetic probes into valuable messengers of the fleeting medium. Thermal dilepton spectra have the potential to offer a direct access to the thermal property of the fireball. Direct photon observables are intimately connected with the dynamical evolution of the bulk medium ,and can be utilized as a viscometer and as a tomographic probe. Electromagnetic probes should continue to play an active role in the future RHIC (BES) program phase II, and in experiments at FAIR. 

Yesterday's puzzle stimulated today's efforts and will become tomorrow's background. The recent theoretical work has increased the relative weight of late hadronic photon emission in the direct photon signal. This produces a larger soft photon emission, accompanied by larger direct photon anisotropic flow coefficients. The net effect is to greatly reduced the tension between theory and the experimental measurements. Indeed, the model-data comparisons have transited from a qualitative description to a more quantitative extraction of the photon emission rates and and a more precise determination of the medium properties. 

Current theoretical calculations are limited by uncertainties linked to non-perturbative physics; the  input from experiments can guide and inform  future  calculations. 
Firstly, theory needs measurements of the low momentum ($p_T < 1$ GeV) photon spectra in pp collisions at the top RHIC and LHC energies. They can reduce the (considerable) theoretical uncertainties in extrapolating the pQCD prompt source to low $p_T$, and provide a reliable baseline for measurements in all  larger collision systems. Secondly, a reduction of the systematic uncertainties in the current direct photon measurements can set stronger constraints on our current knowledge of the photon emission rates and on the dynamical evolution of the collision systems.  Finally, it will be useful and even important  to have direct photon and dilepton measurements in various collisions systems, e.g. (p, d, $^3$He, Al, Cu)+Au, Au+Au and Pb+Pb collisions at different energies: every collision system has its unique space-time evolution structure. A collection of collision systems will provide the important and complementary diagnostic measurements necessary  to probe hot and dense nuclear matter in the different regions of the QCD phase diagram. 

\medskip
{\noindent \bf Acknowledgement:} I thank Gabriel Denicol, Charles Gale, Sangyong Jeon, and Jean-Fran\c{c}ois Paquet for their collaboration, and Elena Bratkovskaya, Wolfgang Cassing, Olena Linnyk, Hendrik van Hees, Min He, and Ralf Rapp for providing the theoretical calculations in Fig.~\ref{fig4}, and for useful discussions. This work was supported by the Natural Sciences and Engineering Research
Council of Canada. 





\bibliographystyle{elsarticle-num}
\bibliography{Shen_C}







\end{document}